\documentclass[aps,reprint,groupedaddress]{revtex4-2}
\usepackage{amsmath}
\usepackage{amssymb}
\usepackage{graphicx}
\usepackage{siunitx}
\usepackage{epstopdf}
\usepackage{relsize}
\usepackage{scalerel}
\usepackage[normalem]{ulem}
\usepackage{esint}

\begin{document}
\title{A numerical approach to levitated superconductors and its application to a superconducting cylinder in a quadrupole field}
\author{J. Hofer}
\email[]{joachim.hofer@univie.ac.at}
\affiliation{University of Vienna, Faculty of Physics, Vienna Center for Quantum Science and Technology (VCQ), A-1090 Vienna, Austria}
\affiliation{Institute for Quantum Optics and Quantum Information (IQOQI), Austrian Academy of Sciences, A-1090 Vienna, Austria}

\date{\today}

\begin{abstract}
Magnetically levitated superconductors in the Meissner state can be utilized as micro-mechanical oscillators with large mass, high quality factors and long coherence times. In previous works analytical solutions for the magnetic field distribution around a superconducting sphere in a quadrupole field have been found and used to derive the trap parameters, while non-spherical geometries have only been investigated in a few idealized cases. However, superconductors of almost arbitrary shape can be used as levitators in a magnetic trap and, as the trap's properties depend strongly on the superconductors shape, allow for a wider parameter regime to be accessed. Finite element models are suitable to obtain the field distribution around arbitrarily shaped superconductors in arbitrary fields, but have not yet been used widely in the context of levitated superconductors.  
Here we present a simple numerical model for this purpose and use it to calculate the field distribution around cylindrical superconductors in a quadrupole field and to evaluate the trap parameters. We find that the cylindrical shape, compared to spherical levitators, allows for substantially higher trap frequencies and coupling strengths. This in turn reduces the demands on vibration isolation and significantly eases the requirements for feedback cooling to the ground state. The numerical model is provided as supplemental material and can easily be adapted to various geometries and trap fields.
\end{abstract}

\maketitle

\section{Introduction}
Magnetically levitated superconductors are already well established as accelerations sensors, most notably in the superconducting gravimeter \cite{Goodkind1999}, which uses centimeter sized hollow spheres coated with Niobium.
More recent proposals \cite{Isart2012,Cirio2012} and experimental advances \cite{Vinante2020, Hofer2023} have shown the potential of using much smaller magnetically levitated particles for approaching the quantum regime with massive, micrometer-sized particles. We believe that the most promising approach is the levitation of a type-I superconductor in the field produced by persistent currents, as this avoids non-fundamental sources of dissipation and decoherence which are present in other levitation schemes \cite{Hofer2023}. For spherical superconductors analytical solutions for the magnetic field distribution, which allow the accurate prediction of the magnetic trap's properties, exist \cite{Hofer2019} and are in good agreement with experimental results \cite{Hofer2023}. For non-spherical superconductors analytical solutions have been found for a few idealized cases, in particular thin rings \cite{Navau2021} and ellipsoids \cite{Navau2024}, and the results indicate that the trap parameters are strongly dependent on the shape of the superconductor and that non-spherical superconductors might allow access to trap parameters that are not attainable with spheres. Furthermore, non-spherical particles possess librational degrees of freedom, which exhibit properties that are not present for the center-of-mass motion \cite{Stickler2021}.
Here we present a numerical model for calculating the field distribution and resulting trap properties for a cylindrically shaped superconductor in a quadrupole field. The simulations are written in \textsc{matlab} and use \textsc{comsol multiphysics} for the numerical evaluation. The model is resource friendly and can be run on a typical personal computer. All files are provided as supplementary material and can easily be modified to evaluate various superconducting geometries (e.g. dumbbells, cuboids). We test the numerical model by computing the field distribution for a spherical superconductor and comparing it to the analytical values, details on this are provided in the appendix. 

\section{Model}
\begin{figure}[htbp!]
\includegraphics[width=0.95\columnwidth]{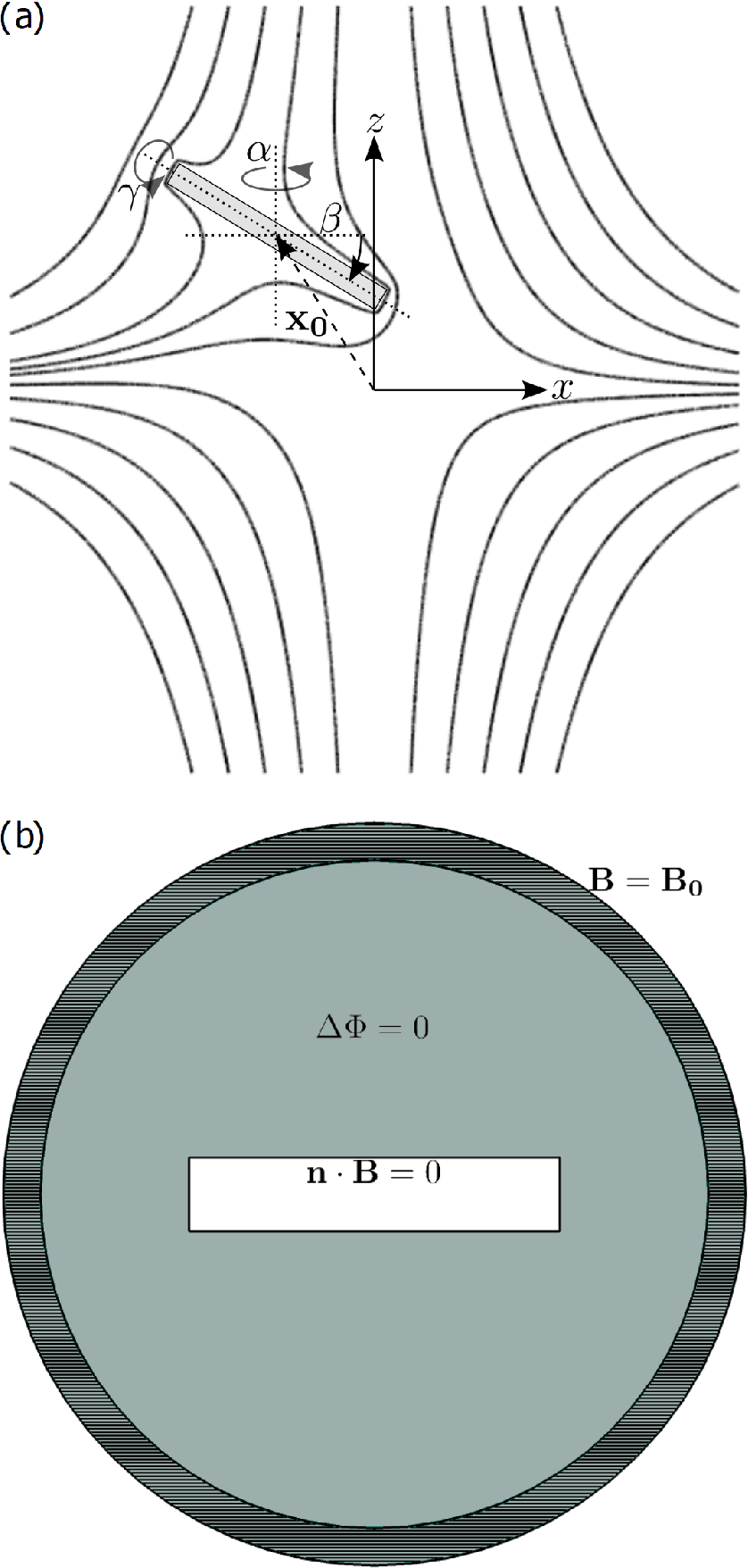}
\caption{\label{fig1} (a)Sketch of the system. The quadrupole  field is depicted by its field lines in the $xz$-plane. (b)Geometrical setup of the \textsc{COMSOL} simulation. In the spherical simulation domain (grey) the equation $\Delta\Phi=0$ is solved numerically, with the boundary conditions being set on the inner and outer boundaries of the domain. In the outer shell of the simulation domain (striped pattern) distances between mesh elements are scaled.}
\end{figure}

We consider a quadrupole trap, such that the applied field is of the form $\mathbf{B_{0}}(\mathbf{x}) = \left(b_x x, b_y y, b_z z\right)$ and, without loss of generality, assume $|b_{x}|<|b_{y}|<|b_{z}|$.  
Since the $b_i$ are related by $b_{z} = - (b_{x} + b_{y})$ (from $\nabla\mathbf{B_{0}}=0$) we can choose a parameter $\epsilon$ with $0\leq\epsilon<1$ such that 
\begin{equation}
\label{eq1}
\mathbf{B_{0}}(\mathbf{x}) = \frac{1}{2}b_z\left((1-\epsilon)x, (1+\epsilon)y, -2 z\right).
\end{equation}
We consider only the case where the cylinder's radius $R$ and height $H$ are both much larger than the superconducting penetration depth $\lambda$, s.t. $\lambda\rightarrow 0$ is a good approximation and the magnetic field vanishes inside the cylinder \cite{Hofer2019}. 
Outside of the cylinder we can introduce a scalar potential $\Phi$ such that 
\begin{equation}
\label{eq2}
\mathbf{B}=\mathbf{B_{0}}-\nabla\Phi \;\text{ and }\; \Delta\Phi=0.
\end{equation}
The magnetic field distribution $B$ around the cylinder is determined by 
\begin{equation}
\label{eq3}
\mathbf{B}\xrightarrow{|\mathbf{x}|\rightarrow\infty}\mathbf{B_{0}}
\end{equation} 
and the boundary condition 
\begin{equation}
\label{eq4}
\mathbf{n}\cdot \mathbf{B}=0
\end{equation} 
on the cylinder's surface, where $\mathbf{n}$ is the surface normal vector.
The position and orientation of the cylinder relative to the field are described by Euler angles in the ZYX convention (the capital letters denote the intrinsic axes, co-rotating with the cylinder) and a displacement vector. Starting from a position where the cylinder's c.o.m. is coincident with the origin of the coordinate system and the cylinder is coaxial w.r.t. the $z$-axis, we first rotate the cylinder by an angle $\alpha$ around the $Z$-axis (which initially coincides with the $z$-axis) and then by an angle $\beta$ around the $Y$-axis. Finally we displace the cylinder's c.o.m. by $\mathbf{x_{0}}$ (cf. Fig.~\ref{fig1}a). The third angle, $\gamma$, which corresponds to a rotation around the $X$-axis, is not needed to describe the position, but will be used further below in the context of the librational modes. 
The setup of the \textsc{comsol} simulations is shown in Fig.~\ref{fig1}b. For the simulations we use the intrinsic coordinate system aligned with the cylinder, such that
\[
\mathbf{B_{0}}\rightarrow R^{-1}_{Y}(\beta)R^{-1}_{Z}(\alpha)\mathbf{B_{0}}\left(R_{Z}(\alpha)R_{Y}(\beta)(\mathbf{x}+\mathbf{x_{0}})\right)
\] 
with 
\[
R_{Y}(\beta) = 
\begin{pmatrix}
  cos(\beta) & 0 & sin(\beta) \\
  0 & 1 & 0 \\
	-sin(\beta) & 0 & cos(\beta)
\end{pmatrix}
\]
and
\[
R_{Z}(\alpha) =
\begin{pmatrix}
  cos(\alpha) & -sin(\alpha) & 0 \\
	sin(\alpha) & cos(\alpha) & 0 \\
	0 & 0 & 1 
\end{pmatrix}
. 
\]
The cylinder is thus always at the center of the spherical simulation domain and constitutes its inner boundary. This setup allows us to reuse the same mesh for all simulations with the same radius and height, reducing meshing times. In the outermost shell of the simulation domain, the distance between mesh elements is scaled such that the boundary condition is set at a distance much larger than any characteristic scale of the superconductor (effectively at infinity). We then use \textsc{comsol} to numerically solve Eq.~\ref{eq2} for the scalar potential in the simulation domain with the boundary conditions (\ref{eq3}, \ref{eq4}) on the inner and outer boundary, respectively. 

Since the set of equations (\ref{eq1} - \ref{eq4}) is scale invariant both w.r.t. length and the magnetic gradient $b_{z}$, it follows that the solution is (up to rescaling) independent of $b_{z}$ and depends only on the ratio $\frac{H}{R}$, i.e.
\begin{equation}
\label{eq5}
\mathbf{B}\left(b_{z}, R, H, \mathbf{x_{0}}, \mathbf{x}\right)=b_{z}R\,\mathbf{B}\left(1, 1, \frac{H}{R}, \frac{\mathbf{x_{0}}}{R}, \frac{\mathbf{x}}{R}\right).
\end{equation}
It is thus sufficient to solve for a fixed value of $b_{z}$ and e.g. $R$, we subsequently use $b_{z}=\qty{1}{T/m}$ and $R=\qty{1}{m}$. This corresponds to measuring length in units of $R$ and magnetic field strength in units of $b_{z} R$.

Having obtained the field distribution, we can calculate the force and torque acting on the cylinder as
\[
\mathbf{F}=\frac{1}{\mu_0}\oiint dS\,\left[(\mathbf{n}\mathbf{B})\mathbf{B}-\frac{1}{2}\mathbf{n}\mathbf{B}^2\right]
\]
and
\[
\mathbf{T}=\frac{1}{\mu_0}\oiint dS\,\left[\mathbf{x}\times\left((\mathbf{n}\mathbf{B})\mathbf{B}-\frac{1}{2}\mathbf{n}\mathbf{B}^2\right)\right],
\]
where the integration runs over the surface of the cylinder and $\mathbf{n}$ denotes the surface normal vector.

Note that sharp edges as used in this simulation are, in principle, unphysical, as in reality the field can always penetrate any physical superconductor to a length on the order of the penetration depth. Indeed, the maximum surface field on the superconductor diverges when a sharp edge is used (analogous to how the magnetic field strength on the surface of a thin wire tends to infinity when the wire size is decreased). We compare the results to a simulation using rounded edges with a finite radius $R_{edge}$ as well as to a simulation of the full Maxwell-London equations taking into account the penetration depth $\lambda$ of the superconductor and show that the results for these cases converge to the results obtained using sharp edges for $R_{edge}\rightarrow 0$ respectively $\lambda\rightarrow 0$ (more details are provided in the appendix).
The model with sharp edges is thus suitable to investigate the general behavior, in practice we recommend using an edge radius corresponding to experimental parameters. Note that the simulations of the full Maxwell-London equations are computationally much more expensive and impractical to run on a typical personal computer, which is why we chose the approach using the scalar potential for the majority of the simulations.

\section{Equilibrium positions}
\begin{figure}[htbp!]
\includegraphics[width=0.95\columnwidth]{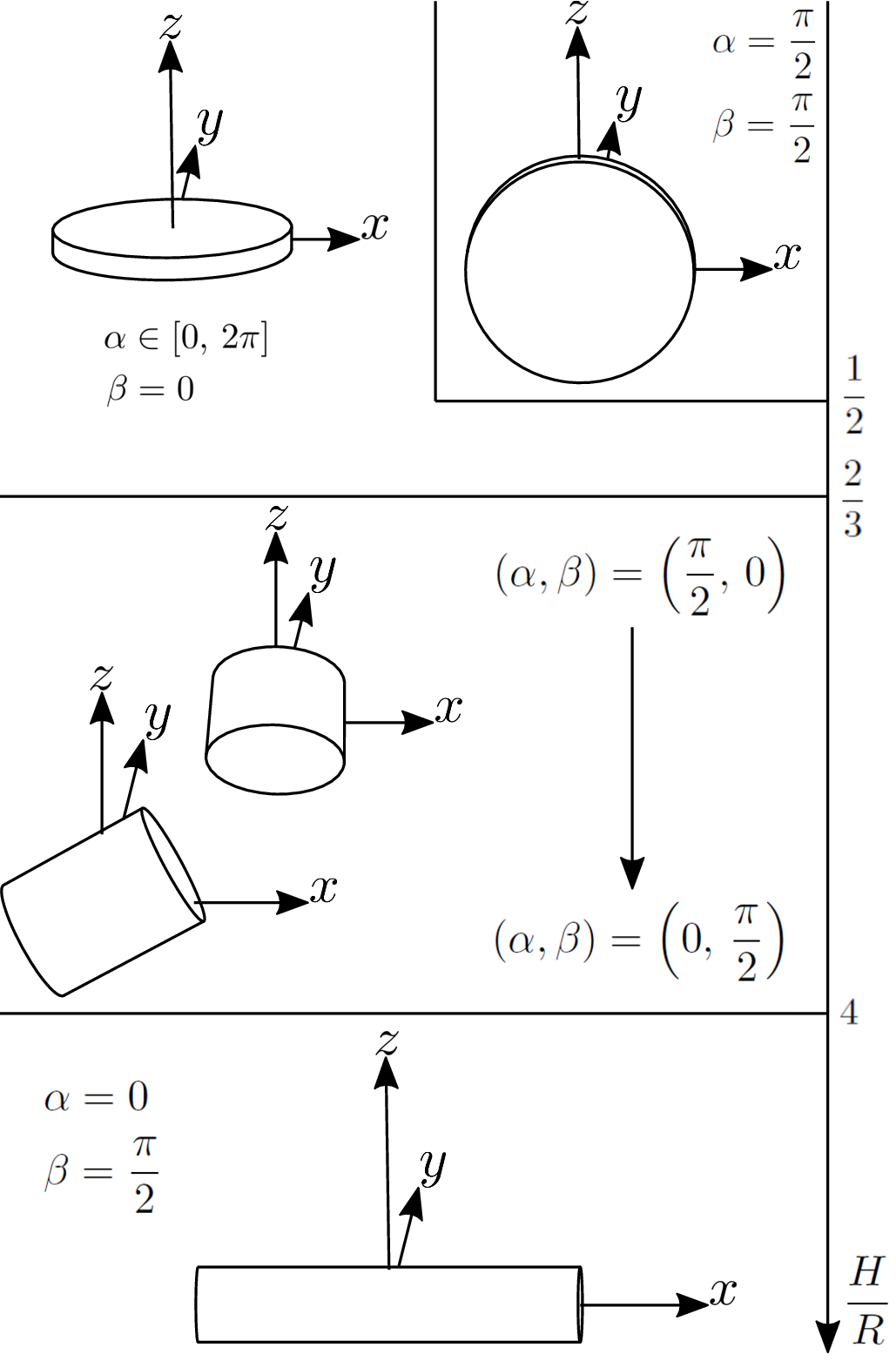}
\caption{\label{fig2} Schematic overview of the different stable trap positions for a superconducting cylinder in a quadrupole trap. For $\epsilon=0$ the cylinder is free to rotate around the $z$-axis and the equilibrium positions is determined solely by $\beta$. Note that the orientation $\alpha=0, \beta=\frac{\pi}{2}$ also results in a stable equilibrium for very small $\frac{H}{R}$ (approximately $\frac{H}{R}<\frac{1}{3}\epsilon$), this is not explicitly shown in the figure. The values for the transition between the different stable orientations shown on the right-hand-side are approximate for $\epsilon = 0$.}
\end{figure}

We show an overview of the stable equilibrium positions in Fig.~\ref{fig2}. 
The c.o.m. position is always $\mathbf{x_{0}}=0$, so the equilibrium positions differ only in $\alpha$ and $\beta$. For simplicity we have chosen not to include the gravitational force into the analysis, which would result in a shift of the rest position along the axis of gravity.
Due to the mirror symmetry of the applied field (w.r.t. the $xy$, $xz$ and $yz$ planes) the mirrored configurations also correspond to stable equilibria.  
Note that for $\epsilon=0$ the cylinder is not constrained at all w.r.t. rotations around the $z$-axis (as the field is radially symmetric), i.e. in that case $\alpha$ can take any value and the equilibria are determined only by the value of $\beta$.

\begin{figure}[htbp!]
\includegraphics[width=0.95\columnwidth]{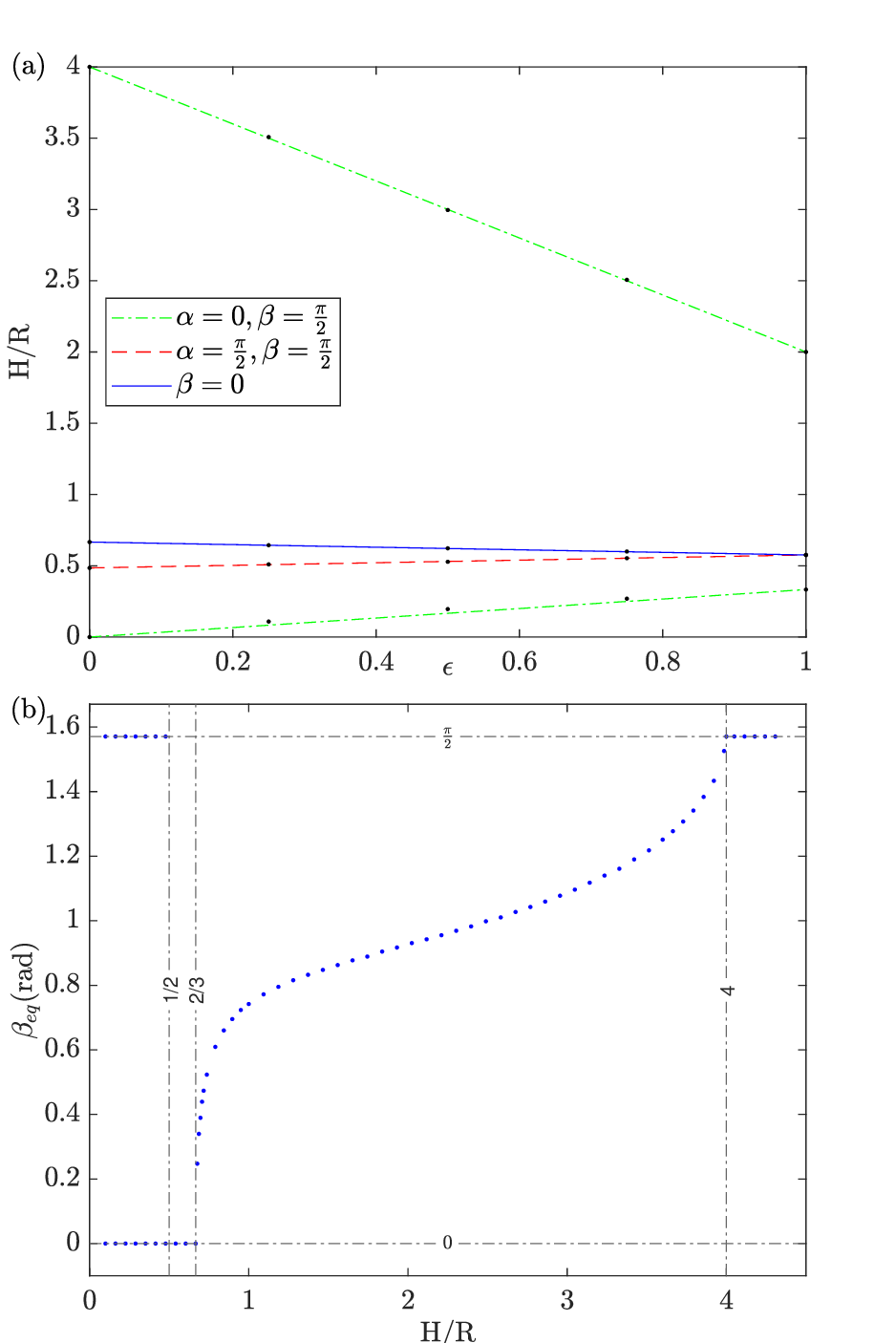}
\caption{\label{fig9} (a)$\epsilon$-dependence of the transition points between the stable equilibria, as described in the text. (b)For $\epsilon = 0$ the equilibrium position of the cylinder is determined solely by $\beta$. The transition points are marked by the vertical lines. For small $\frac{H}{R}$ two stable equilibria exist.}
\end{figure}

\begin{figure}[htbp!]
\includegraphics[width=0.95\columnwidth]{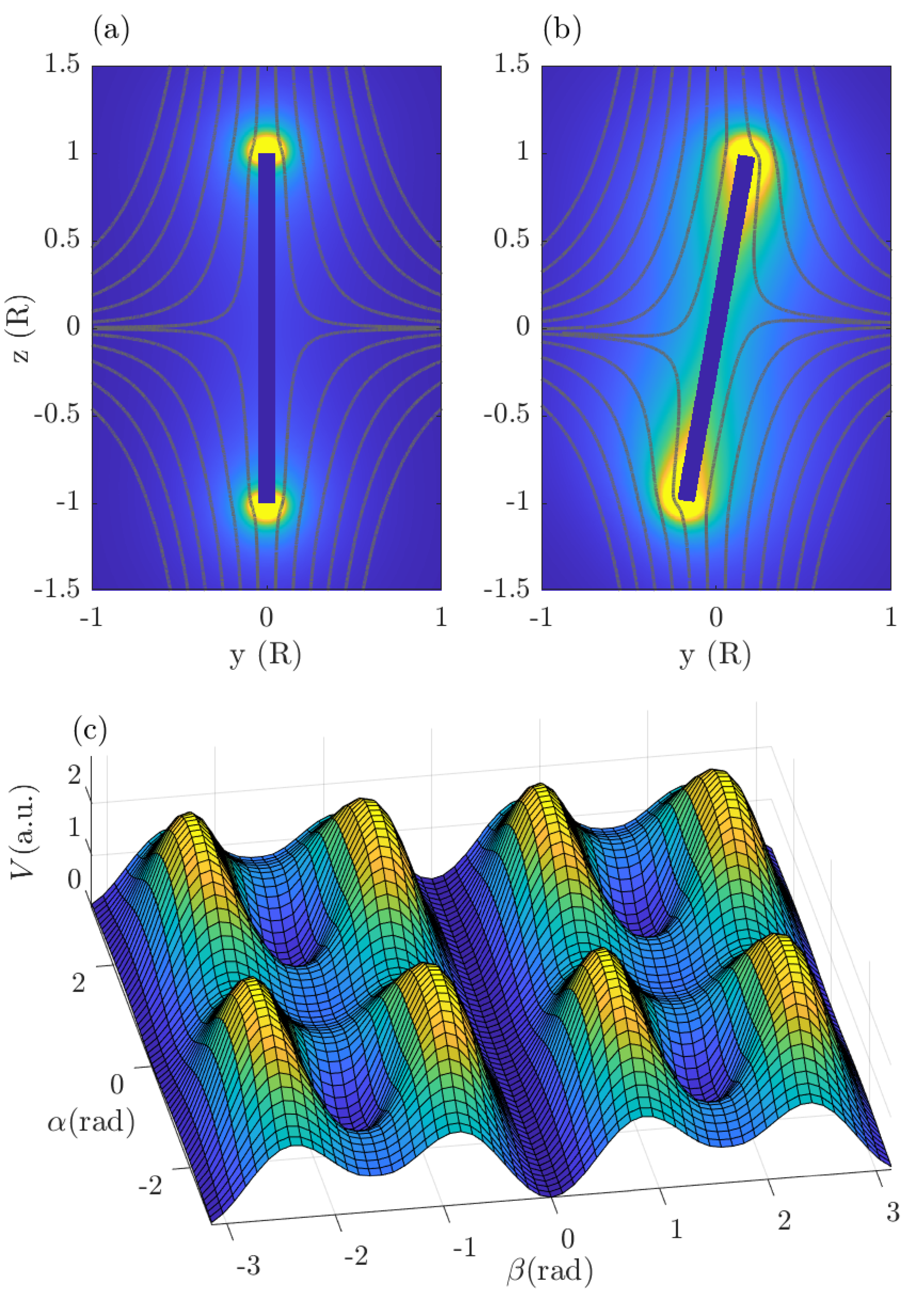}
\caption{\label{fig8} (a,b)Comparison between a cylinder with $H/R=0.1$ in (a) the equilibrium position $\alpha = \frac{\pi}{2}, \beta = \frac{\pi}{2}$ and (b) slightly tilted away from the equilibrium position. The field lines of $\mathbf{B}$ in the $yz$-plane as well as the field norm of the induced field $\mathbf{B}-\mathbf{B_{0}}$ are displayed, brighter colors correspond to higher fields. (c)Magnetic energy w.r.t. $\alpha,\beta$ for a cylinder with $H/R=0.1$ and $\epsilon=0.75$. The stable equilibria correspond to the local minima of the energy.}
\end{figure}

For small $\frac{H}{R}$ a stable orientation is given by $\beta = 0$, while for large $\frac{H}{R}$ the stable orientation is $\alpha = 0, \beta = \frac{\pi}{2}$. In between the orientation changes continuously (cf. Fig.~\ref{fig9}b) from $\alpha = \frac{\pi}{2}, \beta = 0$ to $\alpha = 0, \beta = \frac{\pi}{2}$. 
The exact ratio, at which the transition between the different equilibrium positions occurs, depends on $\epsilon$ (Fig.~\ref{fig9}a), the values stated on the right-hand-side of Fig.~\ref{fig2} are approximate for $\epsilon = 0$.    
Note that for $\frac{H}{R}\lessapprox\frac{1}{2}$ a second equilibrium position exists for $\alpha = \frac{\pi}{2}, \beta = \frac{\pi}{2}$ and for $\frac{H}{R}\lessapprox\frac{1}{3}\epsilon$ there is a third equilibrium at $\alpha = 0, \beta = \frac{\pi}{2}$. This might seem counter-intuitive, since the field gradient is steepest along the z-axis. However, some intuition can be gained from thinking about the way the original field is disturbed by the presence of the superconductor. When the cylinder's flat edges are parallel to the $xz$ or $yz$-plane, the field disturbance is comparatively small (cf. Fig.~\ref{fig8}a). However, if the cylinder is tilted slightly, the disturbance is much larger (cf. Fig.~\ref{fig8}b). It thus makes sense that these orientations result in local energy minima (cf. Fig.~\ref{fig8}c), as the energy is proportional to the volume integral over the squared magnetic field norm. 

\begin{figure}[htbp!]
\includegraphics[width=0.95\columnwidth]{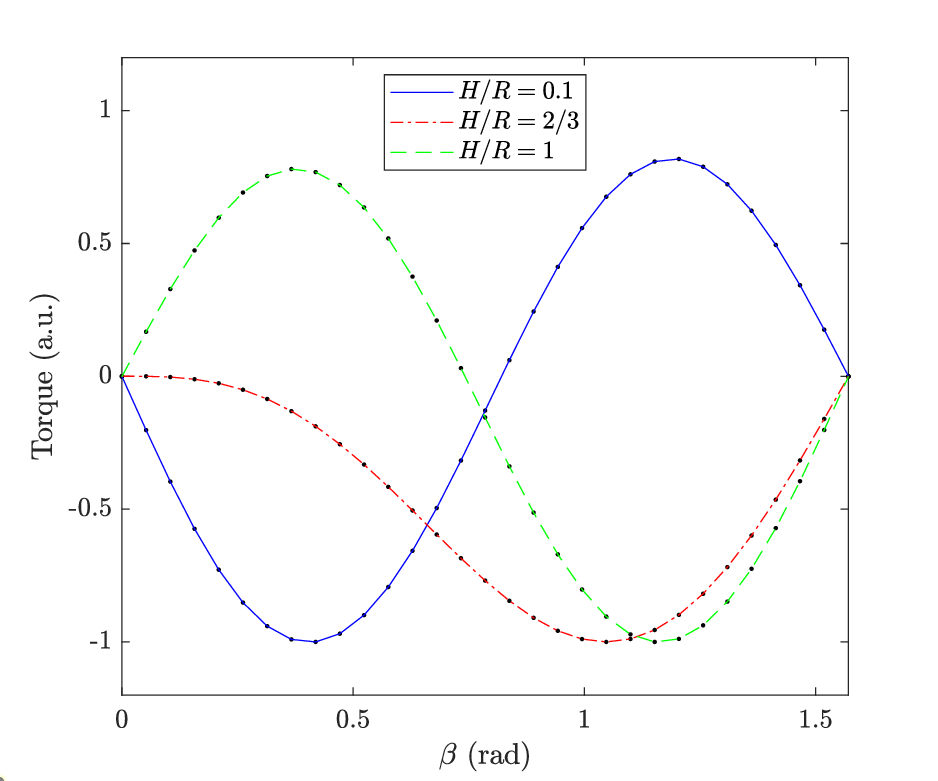}
\caption{\label{fig3} Torque on the cylinder for various aspect ratios and $\alpha =\frac{\pi}{2},\,\epsilon = 0$. For $\frac{H}{R}=\frac{2}{3}$ the harmonic potential vanishes. Each curve was scaled for visibility.}
\end{figure}

In Fig.~\ref{fig3} we plot the torque on the cylinder as a function of the angle $\beta$ for various aspect ratios and $\alpha =\frac{\pi}{2},\,\epsilon = 0$. The linear component of the torque around the rest positions vanishes at the transition points (e.g. $\frac{H}{R}=\frac{2}{3}$), so in this case there is no harmonic trapping potential. 
In the subsequent sections we will focus on the case  $\frac{H}{R}\lessapprox\frac{2}{3}, \beta = 0$, as this position results in the highest trapping frequency for the center-of-mass motion along the $z$-axis as well as the highest ratio of coupling strength to mass, which is relevant for feedback cooling.

\section{Trap frequencies}
For small displacements around the equilibrium position $\beta = 0$ the motion of the cylinder can be characterized by five normal modes, three for the center of mass motion and two librational modes corresponding to rotations around the axes $x$ and $y$. For simplicity we restrict ourselves to the case where the cylinder does not rotate around its symmetry axis ($\dot{\alpha}=0$) and, without loss of generality, set $\alpha=0$ (i.e. $\beta$ and $\gamma$ characterize the librational modes). 

\begin{figure}[htbp!]
\includegraphics[width=0.95\columnwidth]{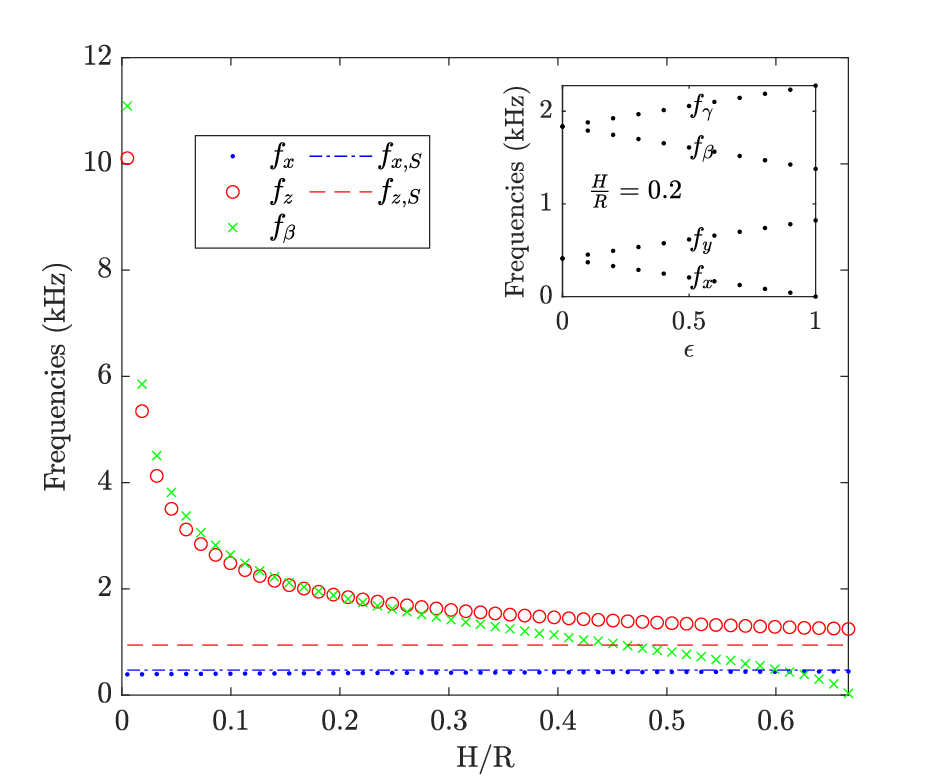}
\caption{\label{fig4} Frequencies for a superconducting niobium cylinder in a quadrupole field with $\epsilon=0$ and $b_{z}=\qty{500}{T/m}$, the dashed and dotted lines show the corresponding values for a superconducting sphere. The inset shows the dependence of the frequencies on $\epsilon$, using $\frac{H}{R}=2$.}
\end{figure}

In Fig.~\ref{fig4}a we plot the frequencies $f_{x}, f_{z}, f_{\alpha}$ for $\epsilon = 0$, in which case the $x$- and $y$-mode as well as the $\beta$- and $\gamma$-mode are degenerate. The inset shows the dependence of these modes on epsilon for $\frac{H}{R}=\num{0.2}$. To plot relatable numbers we have assumed a density of \qty{8570}{kg/m^3} (corresponding to the density of Niobium) and a field gradient $b_{z}=\qty{500}{T/m}$. For comparison, the dashed line displays the frequencies $f_{x,S}, f_{z,S}$ for a superconducting sphere \cite{Hofer2019} using the same parameters. 
Note that the center-of-mass frequency for the $z$-mode is higher than the corresponding frequency for a sphere for all aspect ratios, and can be substantially higher for small aspect ratios. 
It appears that $f_{z}$ could in principle be made arbitrarily high for small aspect ratios, so in practice the limit will likely be the rigidity of the material (or, alternatively, when $H$ approaches the London penetration depth). 

Note that the lateral frequency $f_x$ decreases slowly with decreasing $\frac{H}{R}$. This was contrary to our initial expectations, as we assumed the behavior would be similar to a sphere - a linear dependence of the spring constant on $H$ and thus $f_{x}$ independent of the aspect ratio. We do not have an intuitive explanation for this result, but we double-checked that it is not an artifact of the numerical approach (in particular we ascertained that the values of $f_{x}$ converge with increasing mesh resolution and that the values are independent of the settings of the iterative solver). 

\section{Flux and coupling}
\begin{figure}[htbp!]
\includegraphics[width=0.95\columnwidth]{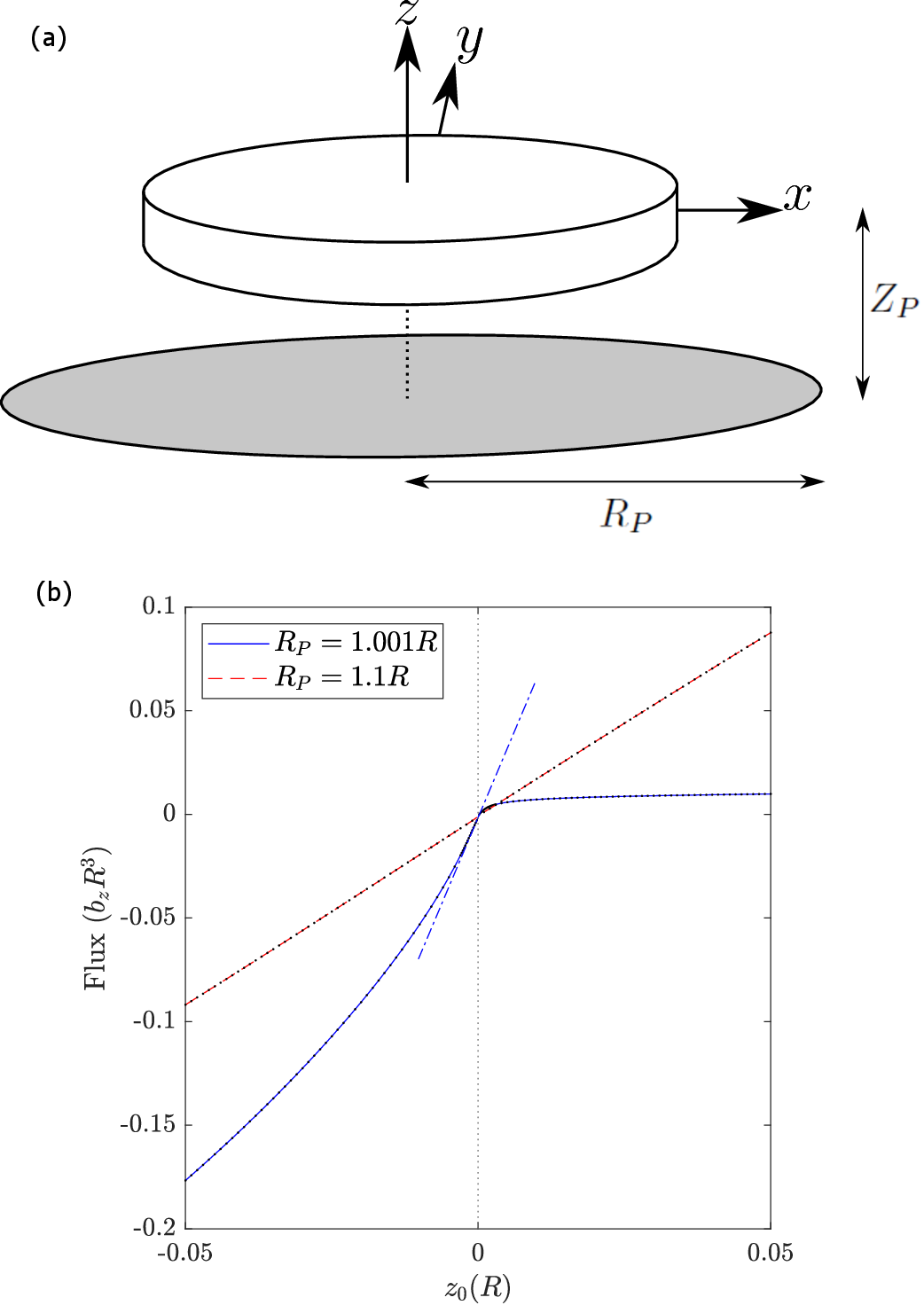}
\caption{\label{fig5} (a)Schematic of the trapped cylinder with a coaxial pickup loop. (b)Flux in a pickup loop with $Z_{P}=\frac{H}{2}$ as a function of particle displacement $z_{0}$. Both curves have been offset s.t. the flux for $z_{0} = 0$ is zero. When the pickup loop is further away from the edge of the cylinder (dashed line), the dependence of the flux on the displacement is linear to a good approximation. When $R_{P}$ is close to $R$ the coupling (corresponding to the slope of the dash-dotted line) increases, but becomes increasingly non-linear.}
\end{figure}

Since the local magnetic field distribution depends on the position and orientation of the superconductor, changes in orientation and position of the levitated cylinder can be read-out by a measuring the change in magnetic flux $\phi$ through a suitable area ("pickup loop") close to the levitation point. For simplicity, we restrict ourselves to evaluating the flux through a planar circular pickup loop coaxial with the $z$-axis. The flux and coupling then depend on only two parameters, the radius of the pickup loop $R_{P}$ and the height above the trap center $Z_{P}$ (cf. Fig.~\ref{fig5}a).
For small oscillations around the equilibrium position and a suitable position of the pickup loop the flux depends linearly on the cylinder's position and can be characterized by a coupling strength $\eta_{z}=\partial_{z}\phi(\mathbf{x_{0}})$. For the numerical evaluation of the coupling we use a displacement of \qty{2.5e-6}{R}, i.e.
\[
\eta_{z} = \frac{\phi(z_{0}=\qty{2.5e-6}{m})-\phi(z_{0}=\qty{-2.5e-6}{m})}{\qty{5e-6}{m}}.
\]
Note that, for the same reasons discussed further above in the context of the maximum surface field, the coupling diverges when the pickup loop approaches the edges of the cylinder and becomes increasingly non-linear (cf. Fig.~\ref{fig5}b). We have thus excluded a small area around $R_{P} = R, Z_{P} = \frac{1}{2}H$ from the evaluation, such that the minimal distance between the edge and the pickup loop is approximately \qty{1e-3}{R}.

\begin{figure}[htbp!]
\includegraphics[width=0.95\columnwidth]{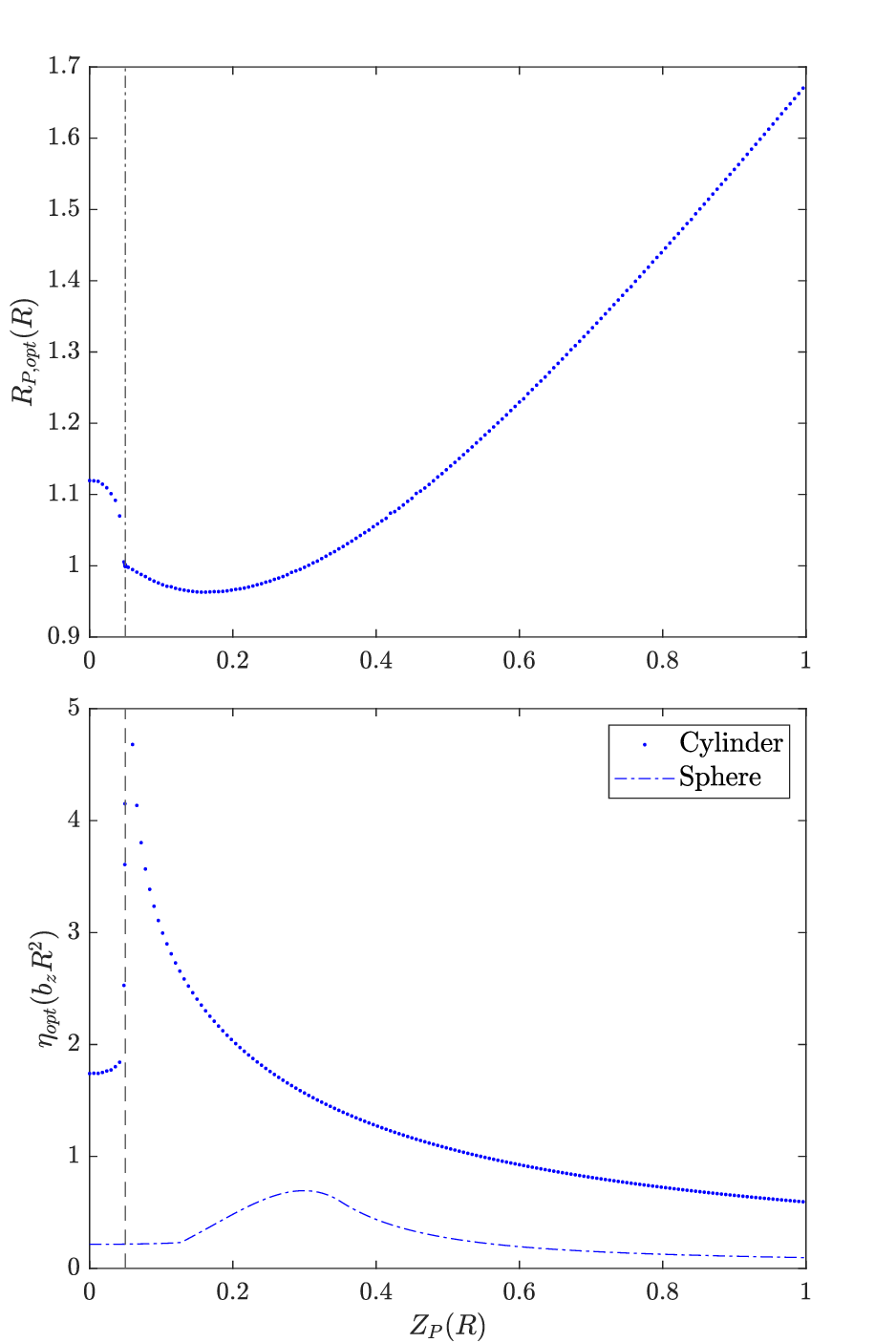}
\caption{\label{fig6} The optimal pickup loop radius (a) and the corresponding coupling (b) as a function of pickup loop position, for $\frac{H}{R}=0.1$. The vertical dashed line denotes $Z_{P}=0.05$, corresponding to the upper edge of the cylinder. The dash-dotted line in (b) shows the optimal coupling to a sphere of the same volume for comparison.}
\end{figure}

For each $Z_{P}$ there is a unique pickup loop radius $R_{P,opt}(Z_{P})$ that results in a maximum coupling strength $\eta_{opt}(Z_{P})$, both of which are plotted in Fig.~\ref{fig6}. Overall, as mentioned above, the coupling is highest when the pickup loop is close to the upper (or lower) edge of the cylinder. We can now compare $\eta_{opt}$ for the cylindrical superconductor to the optimized coupling $\eta_{opt,S}$ for a spherical superconductor \cite{Hofer2023}, where we choose the radius of the sphere such that it has the same volume as the cylinder (because the ratio $\frac{\eta_{z}^{2}}{m}$, where $m$ is the mass of the levitated object, determines the thermal occupation of the $z$-mode under optimal feedback \cite{Hofer2023} and is thus a relevant quantity when comparing different geometries).
We find that the optimal coupling is significantly higher for the cylindrical superconductor (cf. Fig.~\ref{fig6}b) for all pickup loop positions. 
Experimentally the coupling strength can be further increased by using a multi-turn pickup loop and matching the inductance of the pickup loop to that of the sensor \cite{Hofer2023}.

\section{Maximum field}
\begin{figure}[htbp!]
\includegraphics[width=0.95\columnwidth]{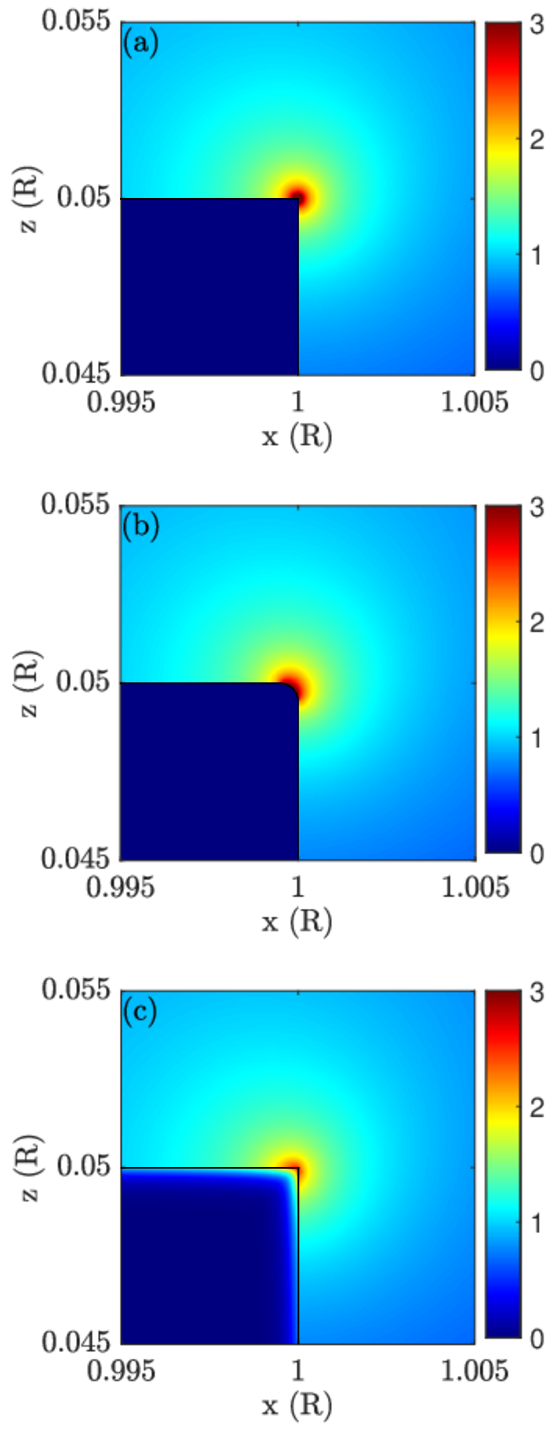}
\caption{\label{fig10} Field norm in the $xz$-plane around at the edge of a cylindrical superconductor with $\frac{H}{R}=0.1$ in the $\beta=0$ equilibrium position for a (a)sharp edge, (b)rounded edge and (c)sharp edge with finite penetration depth. The units on the colorbars are $b_{z}R$.}
\end{figure}

\begin{figure}[htbp!]
\includegraphics[width=0.95\columnwidth]{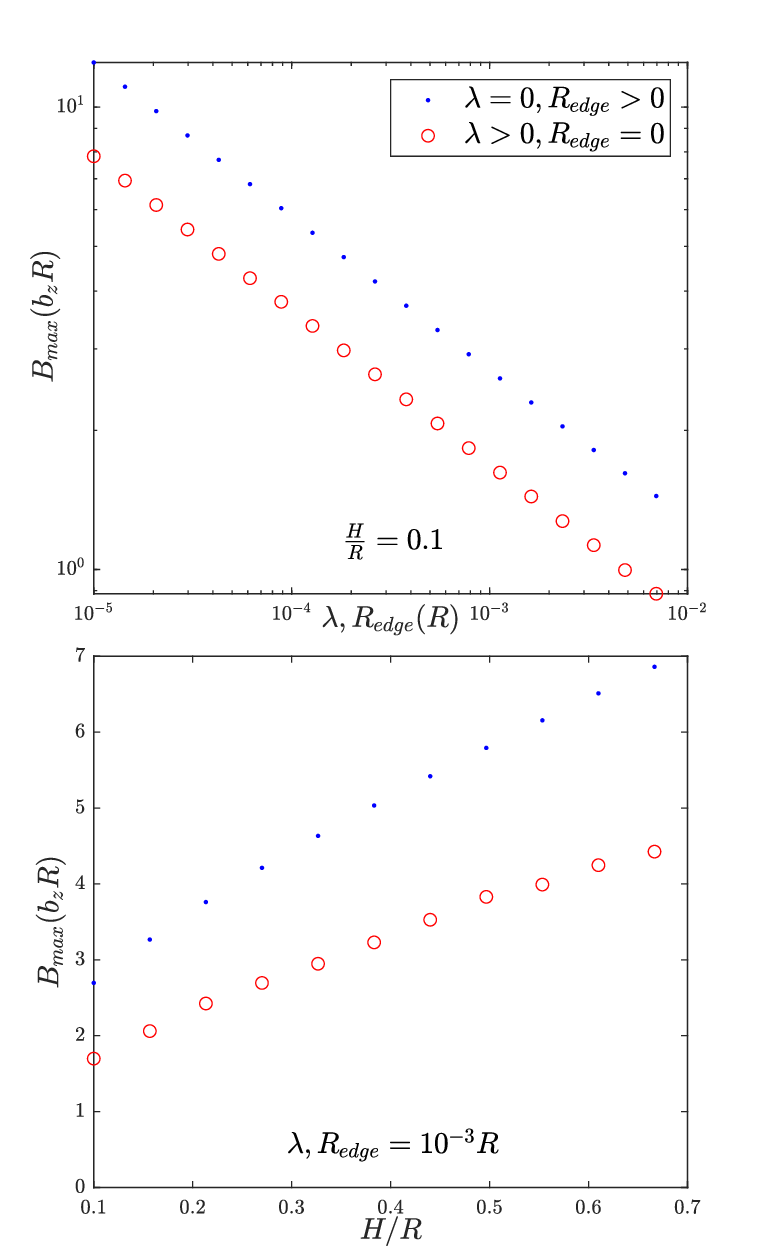}
\caption{\label{fig7} (a)The dependence of the maximum surface field on the edge radius resp. the penetration depth follows a power law with $B_{max}\propto R_{edge}^{-\frac{1}{3}}$ resp. $B_{max}\propto \lambda^{-\frac{1}{3}}$, where the proportionality coefficient is a function of $\frac{H}{R}$, as plotted in (b).}
\end{figure}

The maximum field on the surface of the cylinder is an important parameter to consider, as it has to stay below the (lower) critical field of the superconducting material - otherwise the superconductor would enter either the intermediate state (for type-I superconductors) or the mixed state (for type-II superconductors), which can cause damping of the superconductor's motion due to induced eddy currents or hysteresis. We use the notation $B_{max} = max(|\mathbf{B}|)$, where the maximum is evaluated on the cylinder's surface when it sits at the trap center.
As we've mentioned in the introduction, when the cylinder is modeled with sharp edges, $B_{max}$ diverges (i.e. the numerically evaluated maximum surface field keeps increasing when the resolution of the mesh at the edge is increased). In order to determine $B_{max}$ it is therefore necessary to introduce a rounded edge with a finite edge radius $R_{edge}$ (cf. Fig.~\ref{fig10}).
In Fig.~\ref{fig7}a we plot $B_{max}$ as a function of $R_{edge}$ for $\frac{H}{R}=0.1$ and $\epsilon = 0$, whereas in Fig.~\ref{fig7}b we show the dependence of $B_{max}$ on the aspect ratio.
For comparison we also performed a simulation of the full Maxwell-London equations, where the superconductor has a finite penetration depth $\lambda$, and find that the rounded edge approach overestimates the maximum surface field by a factor of approximately \num{1.6} when using $R_{edge}=\lambda$ (cf. Fig.~\ref{fig7}), likely because the penetration depth does not correspond to a hard boundary and the field can penetrate deeper into the material.
The maximum surface field is typically on the order of $b_{z}R$, so for superconductors on the micrometer scale it is well below typical critical fields (e.g. \qty{80}{mT} for lead or \qty{140}{mT} for the lower critical field of niobium) even for large gradients.

\section{Summary and discussion}
We have built a numerical model in \textsc{comsol multiphysics} to calculate the magnetic field distribution around a superconductor inside a magnetic field. Due to choosing an approach based on the scalar potential (reducing the number of differential equations that need to be solved) the numerical model is not resource-intensive and can be solved on a typical personal computer. The trade-off is that it only applies to geometries large compared to the London penetration depth $\lambda$ (typically on the order of tens of nanometers for type-I superconductors) and that it does not account for flux conservation if the superconductor contains any holes (e.g. in a ring).
We note that there is no added difficulty in numerically solving the full Maxwell-London equations for the vector potential apart from being computationally expensive.

We then applied our model to the special case of a superconducting cylinder in a quadrupole field. The results show that, using a cylinder, it is possible to reach higher trap frequencies and coupling strengths compared to a sphere. Currently, thermal noise and vibrational noise at the trapping frequency are both major obstacles on the path towards ground state cooling a magnetically levitated micrometer-sized superconductor \cite{Hofer2023}. As vibrations decrease with increasing frequency and the thermal occupation of a feedback cooled harmonic oscillator is proportional to $\frac{m}{\eta^{2}}$ \cite{Hofer2023}, using a cylindrical superconductor could significantly relax the requirements for ground state cooling. More generally, these results highlight the importance of the superconductor's geometry in determining the trap parameters and that non-trivial geometries should be investigated, even if an analytical solution is not available. 

Going forward, we plan to introduce non-spherical superconductors in our experiments (cf. \cite{Hofer2023}) to take advantage of these findings. Also, having found that the shape of the superconductor plays a major role in determining the properties of the trap, we will study more complex geometries to identify the geometry that best suits our application. 

The numerical model is available at the Zenodo repository \cite{hofer_2024_10991398}. As it can be easily adapted to different superconducting geometries and trapping fields, we hope that it can serve as a useful tool for other researchers as well to find the optimal experimental arrangement for their purpose without having to rely exclusively on analytically solved models.

\begin{acknowledgments}
I am grateful for discussions with C. Navau Ros, J. Cunill Subiranas and N. J. Bort Soldevila, as well as discussions with P. Asenbaum, M. Aspelmeyer, R. Claessen, J. Hansen, G. Higgins, P. Schmidt and M. Trupke. 
This work was supported by the European Union‘s Horizon 2020 research and innovation programme under grant agreement No 863132 (iQLev), the European Union‘s Horizon Europe 2021-2027 Framework Programme under grant agreement No 101080143 (SuperMeQ), the European Research Council (ERC CoG QLev4G and ERC Synergy QXtreme), and the Austrian Science Fund (FWF) under project F40 (SFB FOQUS).
\end{acknowledgments}

\bibliography{ms}

\appendix

\section{Convergence and comparison to the analytical solution}
The accuracy of the numerical solution typically depends on various parameters, sources of errors include a geometrical error due to discretizing the geometry and an algebraic error depending on the termination criteria for the iterative solver. Ideally these errors can be made arbitrarily small by choosing a sufficiently well resolved mesh as well and a suitable termination criteria for the solver. Testing these errors is usually done by changing the parameters (such as mesh size and distribution, size of the simulation domain and relative size of the infinite element domain, solver tolerance and precision) and observing how the solution changes. If the numerical solution is already close to the actual solution, further refinements in these parameters should not result in large changes in the solution, i.e. the solution should converge to a fixed value w.r.t. these parameters. 
\begin{figure}[htbp!]
\includegraphics[width=0.95\columnwidth]{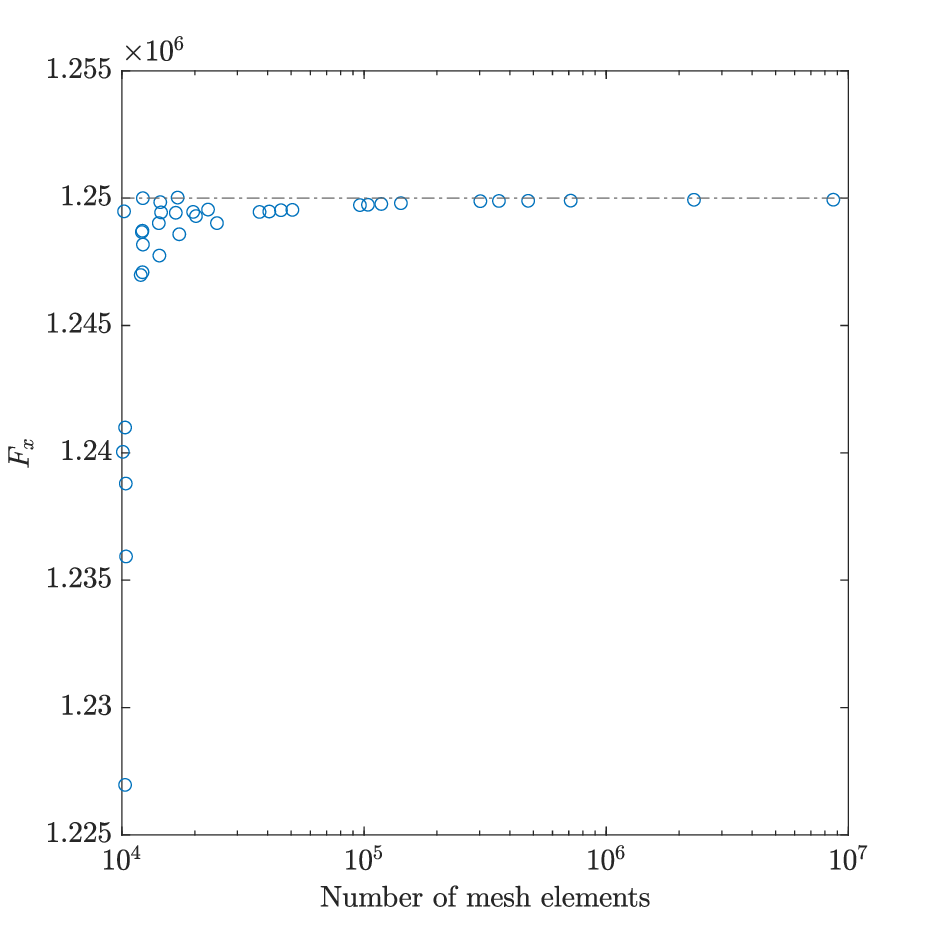}
\caption{\label{figS1} Convergence of $|F_{x}|$ for a spherical superconductor with increasing mesh resolution. The dash-dotted line corresponds to the analytical solution.}
\end{figure}

In our case, we can additionally choose a geometry were the analytical solution is known (spherical superconductors), to verify that the numerical solution converges to the analytical solution for suitable mesh and solver settings. In Fig.~\ref{figS1} we show an example of a convergence plot w.r.t. $|F_{x}|$ (for a sphere radius $R_{S}=\num{1}$ $b_{z}=\qty{1}{T/m}$, $\epsilon=\num{0}$ and $\mathbf{x_0}=(1,0,0)$) and the comparison to the analytical solution of \qty{1.25e6}{N} \cite{Hofer2019}. The numerical solution converges to the analytical solution with a relative error of less than \qty{0.01}{\percent}. 
The same behavior applies to all other evaluated quantities (force components and quantities derived thereof, field distribution, maximum surface field) and all show excellent agreement with the analytical values. 

\section{Dependence of the solutions on the edge radius or penetration depth}
As mentioned in the main text, sharp edges on the cylinder lead to the divergence of the surface field at the edge (and, as a direct consequence, for the coupling strength to a pickup loop aligned with the edge). It is important to check that the results presented above, most of which are derived from the force and torque (and thus from surface integrals involving the magnetic field) are well-behaved. As described in the last section, we first check that the results for force and torque don't depend on the resolution of the mesh for a sufficiently fine resolution, i.e. that the numerical results converges with increasing mesh resolution.  
\begin{figure}[htbp!]
\includegraphics[width=0.95\columnwidth]{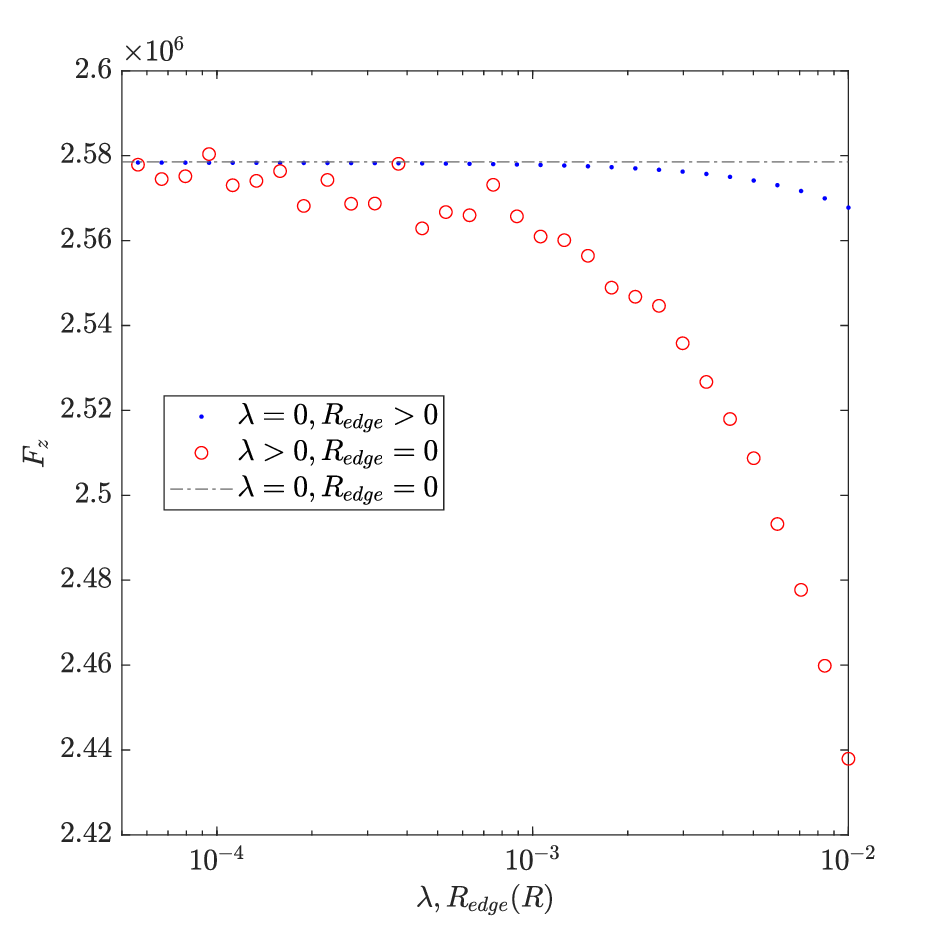}
\caption{\label{figS2} The results of simulations with rounded edges or finite penetration depths converge for $R_{edge}\rightarrow 0$ resp. $\lambda\rightarrow 0$. The dash-dotted line corresponds to the result obtained when using $R_{edge}=0$ and $\lambda=0$.}
\end{figure}
We then performed simulations with a finite edge radius $R_{edge}$ and checked that the results converge for $R_{edge}\rightarrow 0$. Finally we did the same test using a model of the full Maxwell-London equations and a finite penetration depth $\lambda$. This test was only performed for some results, as solving this model on a personal computer with the required accuracy takes a long time. An example is shown in Fig.~\ref{figS2}. 
For the maximum surface field we have already verified that it diverges for $R_{edge}\rightarrow 0$ and $\lambda\rightarrow 0$, this is described in the main text.

\end{document}